\input{aipcheck}

\documentclass[
    ,final            
  ]
  {aipproc}

\layoutstyle{6x9}

\begin{document}

\title{A Decade of Dark Energy: 1998 - 2008}


\author{Ruth A. Daly}{
  address={Department of Physics, Penn State University, Berks Campus,  
P. O. Box 7009, Reading, PA 19610}
}

\begin{abstract}
The years 1998 to 2008 were very exciting years for cosmology.  
It was a pleasure to accept this invitation to describe my contributions
to the development of our knowledge and understanding of 
the universe over the course of the past decade.  
Here, I begin by 
describing some of my work on radio galaxies as a 
modified standard yardstick and go on to describe   
model-independent studies of the accelerating universe and the 
properties of the dark energy. During the course of these studies,
I came upon interesting ways to study the spin and other 
properties of supermassive black holes, 
some of which are briefly mentioned. 
\end{abstract}

\maketitle


\section{The Early Years}
Many scientists have contributed to our knowledge and understanding 
of the accelerating universe and the properties of the dark energy.  
In keeping with the request of the conference organizers, this paper 
only addresses my contributions to this field. 

The years 1997 and 1998 were exciting times for cosmology! I had
been studying powerful classical double radio galaxies since 
the early 1990s \cite{D90}, and had proposed a new method of 
using radio galaxies as a modified standard yardstick in 
1994 \cite{D94}.  This work continued with 
Princeton University PhD thesis students Eddie Guerra, 
Lin Wan, and Greg Wellman, and some of this work included cosmological 
studies (\cite{GD98}, 
\cite{DGW98}, \cite{GDW98}, \cite{GDW00}, 
\cite{DG02}), and studies of outflows from the 
supermassive black holes that power the radio sources 
(\cite{WDW97a}, \cite{WDW97b}, \cite{WD98a}, \cite{WD98b}, 
\cite{WDG00}). The cosmological studies were done in the context of 
two cosmological world models: one that
included non-relativistic matter, a cosmological constant, and space
curvature, and another that included 
 ``quintessence'' with constant equation of
state, non-relativistic matter, and zero space curvature.
Later, radio galaxies were studied in the context of a cosmological
model that included a rolling scalar field, 
non-relativistic matter, 
and zero space curvature \cite{P03}, \cite{D09a}.

The cosmological 
results eventually published by \cite{GDW00} were presented
on January 9, 1998 at the AAS meeting in Washington, D. C.;  
the sample of twenty radio galaxies studied 
is briefly mentioned in AAS Bulletin Abstract 95.04.  These results  
indicated that a cosmological constant provided a good fit
to the radio galaxy data. 
In late 1997, Steve Maran from the AAS press office invited me 
to prepare a press release on cosmological studies with radio
galaxies to be presented on January 8, 1998, and I accepted this invitation. 
The press release \footnote{available at 
{http://www.princeton.edu/pr/news/98/index1.html}
under ``The Ultimate Fate of the Universe'' (1/8/1998) and at  
{http://www.bk.psu.edu/faculty/daly}} 
explains how distant radio galaxies can be used 
to study the expansion history of the universe. For  
a given observed angular size of the radio source, 
a large intrinsic size meant that the coordinate distance
to the source was large, and that 
the universe was accelerating in its expansion, or, in the 
words of the release, ``the expanding universe will continue to 
expand forever, and will expand more and more rapidly as time goes by.'' 
This is explained again later in the release where it states  
``the universe will continue to expand forever and will expand at
a faster and faster rate as time goes by.''

The press release session included supernova results
presented by Adam Riess and Saul Perlmutter, and I had an opportunity
to discuss my conclusions with Adam and Saul in detail.  
They were surprised to hear that a cosmological constant provided
a good fit to the radio galaxy data, which implied the universe
would expand at an ever increasing rate. Each expressed a similar concern,
this concern being the dependence of the result on the cosmological model or 
world view under consideration.
The result that I was reporting on 
was obtained in the context of a cosmological model that allowed for 
non-relativistic matter, a cosmological constant, and space curvature.
If different components were present in the universe, 
would the data still imply that
the universe was accelerating?  In fact, addressing this concern was 
part of the motivation for developing a completely model-independent 
approach to the analysis and interpretation of radio galaxy and 
supernova data (described below).  

Two supernova groups showed that a cosmological constant provides
a good description of the supernova data  
(e.g. \cite{R98}, \cite{P99}).   
As time passed and more data were analyzed,  
it became clear that these two completely independent methods, FRIIb 
radio galaxies and type Ia supernovae, based on totally different types of
sources and source physics, yield very similar results 
\cite{DMG02}, \cite{D05}, \cite{DD03}, \cite{DD04}, 
\cite{DD05}, \cite{DD06}, \cite{D08}. 
This was important because it suggested that
neither method was plagued by unknown systematic errors.  

One of the reasons the radio galaxy method provides 
interesting results with a relatively small number of sources 
is that many of the radio sources are
at relatively high redshift. For example, the highest redshift 
source in either the radio galaxy or supernovae samples is 
the radio galaxy 3C 239 at a redshift of 1.79, which has been
included in the radio galaxy studies since 1998 \cite{GD98}. 
Differences between predictions of various 
cosmological models become large at high redshift, so  
high redshift data points can have a strong impact on results. 

\section{Understanding the Radio Galaxy Method}
The methods of using type Ia supernova and type IIb 
radio galaxies for cosmological
studies are empirically based. It could be empirically demonstrated
that the methods worked well,
but the underlying physical processes were not understood well enough to 
explain why the methods worked so well.  This changed in 2002
for the radio galaxies, when the reason that the radio galaxy
method works so well began to become clear \cite{DG02}.
The radio galaxy method is applied to very 
powerful classical double radio galaxies, such as the radio source
Cygnus A (3C 405). These FRIIb radio galaxies 
are powered by very energetic, highly
collimated outflows from regions very close to a supermassive
black hole located at the center of a galaxy. 
When the collimated outflow impacts the ambient gas, a strong
shock wave forms, and a shock front separates the radio emitting 
material from the ambient gas \cite{D90}.  The physics of strong shocks
is fairly simple and straight-forward, and makes these
systems ideal for cosmological studies. 

Large-scale outflows from supermassive black holes 
are thought
to be powered by the spin energy of the hole 
(e.g. \cite{BZ77}, \cite{R84}, \cite{B90}).  
When \cite{DG02}
cast the radio galaxy method in the language of the Blandford-Znajek
model  to extract the spin energy 
from a rotating black hole \cite{BZ77}, it became clear that the 
outflow from the hole occurs when the strength of the magnetic field 
near the hole reaches a maximum or limiting value.  This value 
can be written as a function of the black hole mass, spin, and 
the radio galaxy model parameter, $\beta$.  When the radio galaxy 
model parameter has one particular value, $\beta = 1.5$, the 
relationship between the magnetic field strength and the properties of
the rotating hole is greatly simplified, and the field strength 
depends only upon the black hole spin.  
Empirical studies by \cite{DG02}
and \cite{D09a} found 
that the value of $\beta$ is very close to 1.5, $\beta \simeq 1.5 \pm 0.15$.
Thus, the reason the radio galaxy model works so well is that 
the outflow from the supermassive black hole is triggered when the 
magnetic field strength reaches a maximum or limiting value 
that depends only upon the black hole spin. 
Interestingly, other models, such as that by \cite{M99},  
have the same functional form as the Blandford-Znajek model \cite{BZ77} 
but with a different constant of proportionality, and the  
results of \cite{DG02} apply to any model with the same functional 
form as the  Blandford-Znajek model.  

\section{The Model- Independent Approach}
From 1998 to 2002 the study of the acceleration of the universe was done 
in the context of particular cosmological world models, 
and the question 
of whether the acceleration of the universe could be studied independent
of a particular cosmological model and independent of a theory of 
gravity captivated my interest. To address this question, I worked to develop 
an assumption-free, or model-independent, 
method of analyzing supernova, radio galaxy, or other data sets that
provide coordinate distances. The method
was proposed in 2002 \cite{D05},  
and, in collaboration with George Djorgovski, 
was developed and applied to supernova
and radio galaxy data sets \cite{DD03}, \cite{DD04}, \cite{DD05}, 
\cite{DD06}, \cite{D08}.  

Assuming only that the Friedmann-Lema\^{i}tre-Robertson-Walker 
(FLRW) line element 
is valid, coordinate distance measurements can be used to obtain
the expansion and acceleration rates of the universe as 
functions of redshift. Coordinate distance measurements
are easily obtained from luminosity distances or angular size
distances to any type of source (e.g. supernovae or radio galaxies). 
The FLRW line element is the most
general metric describing a homogeneous and isotropic four-dimensional 
space-time. 
These determinations of the expansion and acceleration rates of the 
universe are independent of 
a theory of gravity, and independent of the contents of the 
universe (\cite{D05}, \cite{DD03}). 
It was shown by \cite{D08} that 
the zero redshift value of the dimensionless acceleration 
rate of the universe is independent of space curvature,
and that very similar
results are obtained for the dimensionless acceleration rate 
$q(z)$ and the expansion rate of the 
universe $H(z)$ for 
zero and reasonable 
non-zero values of space curvature.  Thus, the model-independent 
method can be applied 
without requiring that space curvature be set equal to zero.  

It was shown by \cite{DD03}, \cite{DD04}, \cite{DD05}, \cite{DD06}, 
and \cite{D08} that 
the universe is accelerating today and was most likely decelerating
in the recent past, 
and this result is independent of a theory of
gravity, of the contents of the universe, and of whether space
curvature is non-zero (for reasonable non-zero values).  

Recent determinations of $H(z)$ and $q(z)$ 
obtained using the model-independent method are compared 
with predictions in
a standard Lambda Cold Dark Matter (LCDM) model in Fig. 1 (the thin 
solid line shows the LCDM prediction). These results  
indicate that 
the LCDM model 
provides a good description of the data to a redshift of about one 
(e.g. \cite{DD03}, \cite{DD04}, \cite{DD05}, \cite{DD06}, 
\cite{D08}). The LCDM model
assumes that General Relativity (GR) is the correct theory of gravity, 
space curvature is equal to zero, and two components contribute to
the current mass-energy density of the universe, a 
cosmological constant and non-relativistic matter  
with 70 \% and 30 \%, respectively, 
of the normalized mean mass-energy density of the universe at the 
current epoch.  As discussed by \cite{D08}, 
a comparison of model-independent determinations of $H(z)$ and $q(z)$
with predictions in the LCDM and other models 
provides a large-scale test of GR. 
Current observations suggest that GR 
provides an accurate description of the data over look back times
of about ten billion years \cite{D08}.  
There is a hint of a deviation of the 
data from predictions in the LCDM model at redshifts of 
about one (\cite{DD04}, \cite{D08}). 

The model-independent approach can be extended to solve for the 
properties of the dark energy as a function of redshift \cite{DD04}, where   
the ``dark energy'' is the name given to whatever is causing the 
universe to accelerate.  Assuming that 
GR is valid on very large length scales, and that space curvature
is zero, supernova and radio galaxy data can be used to solve
for the pressure, energy density, equation of state, and potential
and kinetic energy densities of the dark energy as functions
of redshift (\cite{DD04}, \cite{D08}), as shown in Fig. 2.  
Results obtained using the model-independent approach can provide
valuable information to theorists developing new ideas to
explain the acceleration history of the universe and the properties of the 
dark energy. 
This is complementary to the commonly adopted approach of assuming  
a particular dark energy model and cosmological world model
and solving for best fit model parameters (e.g. \cite{R98}, \cite{P99}, 
\cite{DGW98}, \cite{GDW00}, \cite{DG02}, \cite{P03}, \cite{D09a}). 

In studies of the properties of the dark energy, the equation of
state of the dark energy has surfaced as an important
parameter.  A cosmological constant has an equation
of state that is always equal to $-1$. 
To study the equation of state of the dark energy 
in a model-independent manner, \cite{D08}  
defined a new model-independent function,
called the dark energy indicator.  The dark energy indicator provides
a measure of deviations of the equation of state from $-1$ as a function 
of redshift.  Current data suggest that a value of $w=-1$ provides
a good description of data at redshift less than 1 (see Fig. 2).

\section{Supermassive Black Holes and their Spins}

The radio galaxies described above are powered
by large-scale outflows from the vicinity of supermassive 
black holes. Studies of the properties of a radio galaxy
allow the energy per unit time, known as the ``beam power,''
that is being channeled from the vicinity of the supermassive black hole 
to the large-scale outflow to be determined. 
Studies of the beam power and other source properties provide
important insights and information on these black hole systems
(e.g. \cite{WDG00}, \cite{D95}, \cite{O09}). 

For example, the beam power can be combined with the radio galaxy model 
parameter $\beta$ to solve for the total energy that will be 
channeled away from the vicinity of the supermassive black hole
over the full lifetime of the outflow (e.g. \cite{WDG00}, \cite{DG02}, 
\cite{O09}). The total energy 
of the outflow can be combined with the black hole
mass to obtain a lower bound on the spin of the supermassive
black hole \cite{D09b}, assuming only that the highly collimated
outflow is powered by the spin energy of the supermassive black hole. 
This is one of the very few direct indications of the spin of
supermassive black holes known at present. 
The ratio of the total outflow energy to the black hole mass 
appears to be constant for these black hole systems. 
This ratio provides an important diagnostic of the physical
state of the black hole system at the time the outflow is
generated, and the results of \cite{D09b} indicate that 
each system is in a similar physical state when 
the outflow is triggered. Thus, these studies provide insights 
into the physical conditions of  
supermassive black holes systems and their state
at the time powerful outflows are generated. 

\begin{figure}
\begin{minipage}[t]{7.75cm}
\begin{center}
\includegraphics[width=7.75cm,clip]{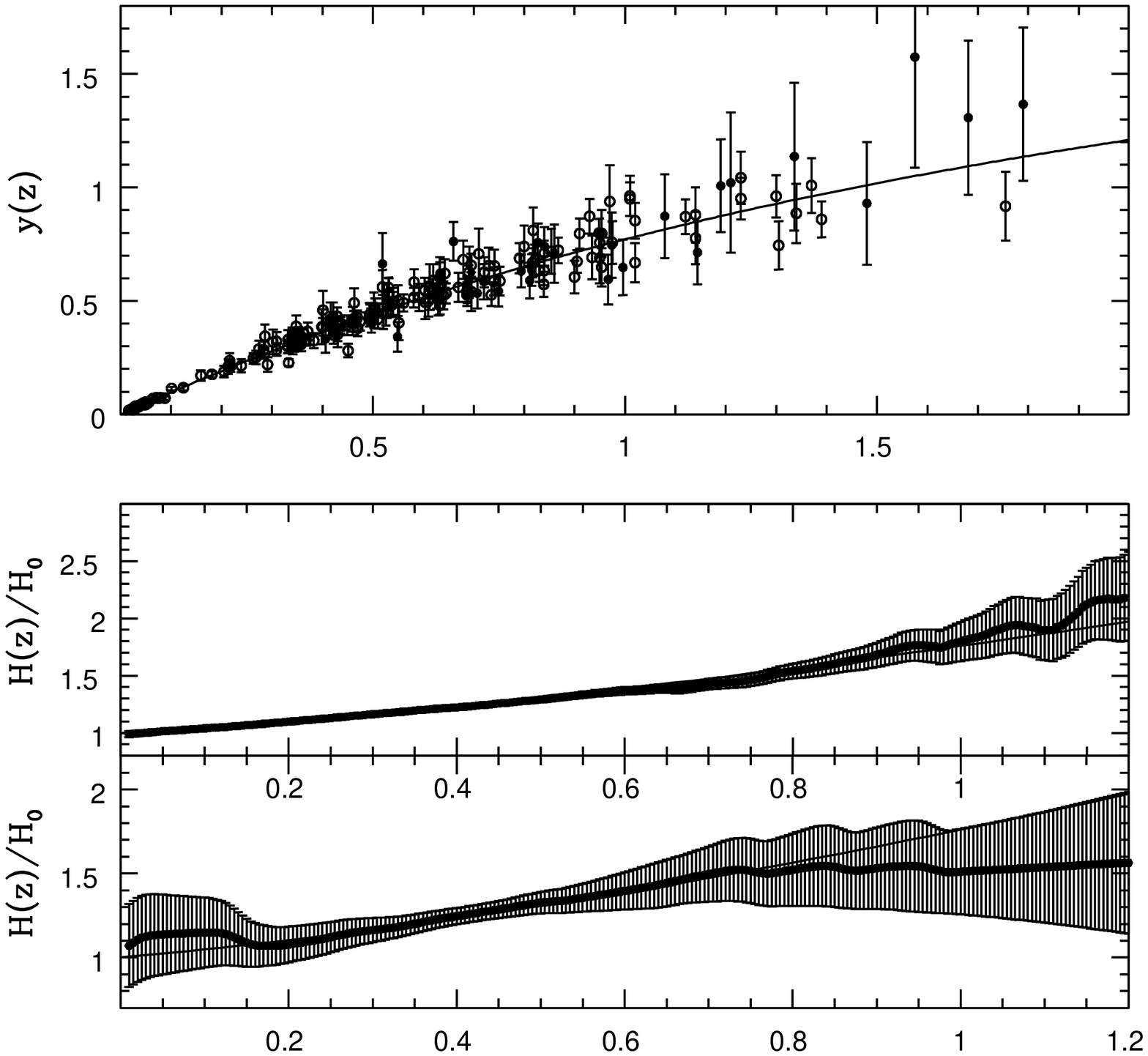}
\end{center}
\end{minipage}
\hfill
\begin{minipage}[t]{7.75cm}
\begin{center}
\includegraphics[width=7.75cm,clip]{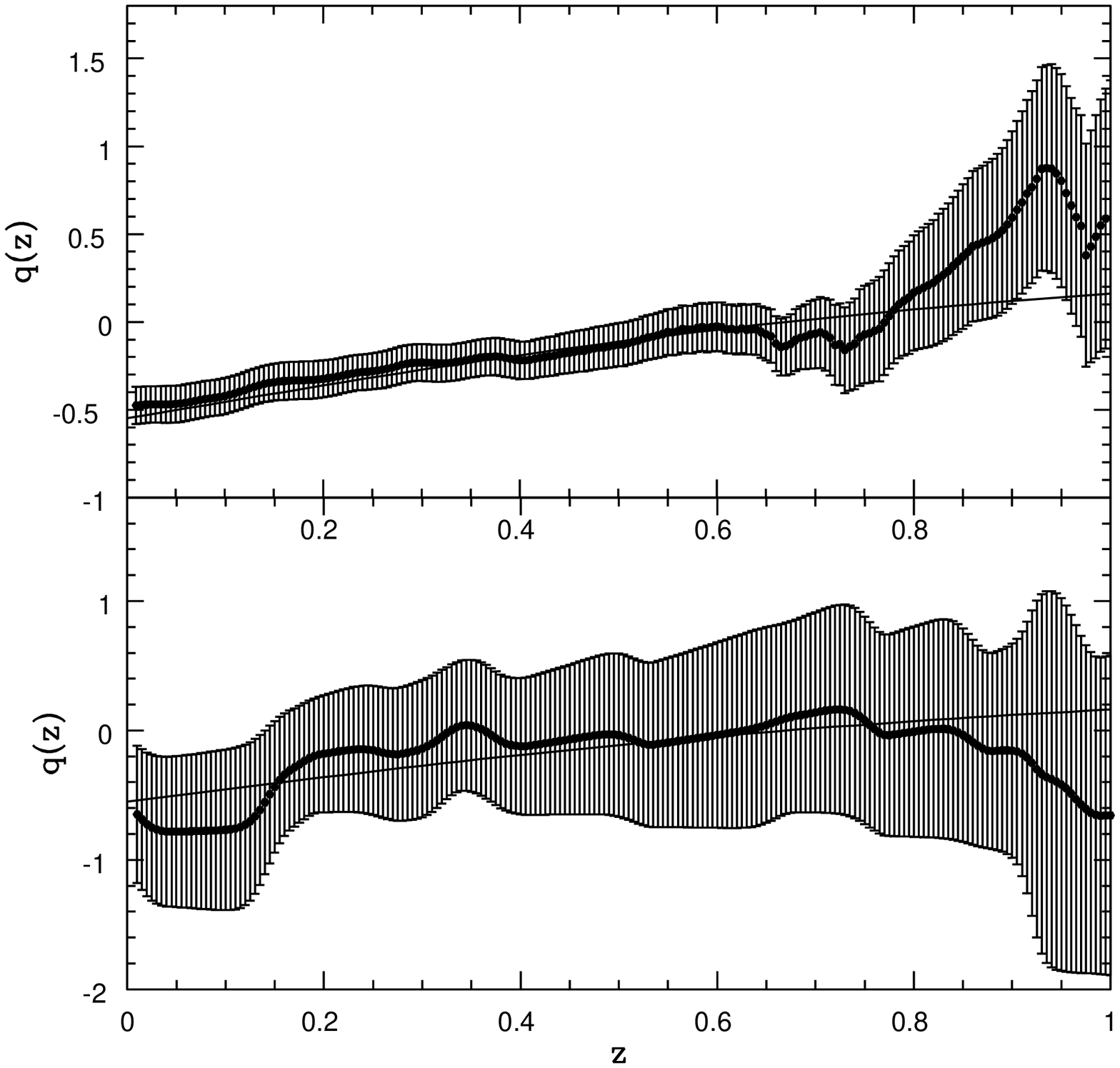}
\caption{\label{fig1}  Dimensionless coordinate distances 
(filled symbols indicate radio galaxies and open symbols indicate
supernovae),  
$H(z)$ for supernovae (middle left panel)
and radio galaxies (bottom left panel), 
and $q(z)$ for supernovae (top right panel)
and radio galaxies (bottom right panel); from Daly et al. (2008). }
\end{center}
\end{minipage}
\end{figure}

\begin{figure}
\begin{minipage}[t]{7.75cm}
\begin{center}
\includegraphics[width=7.75cm,clip]{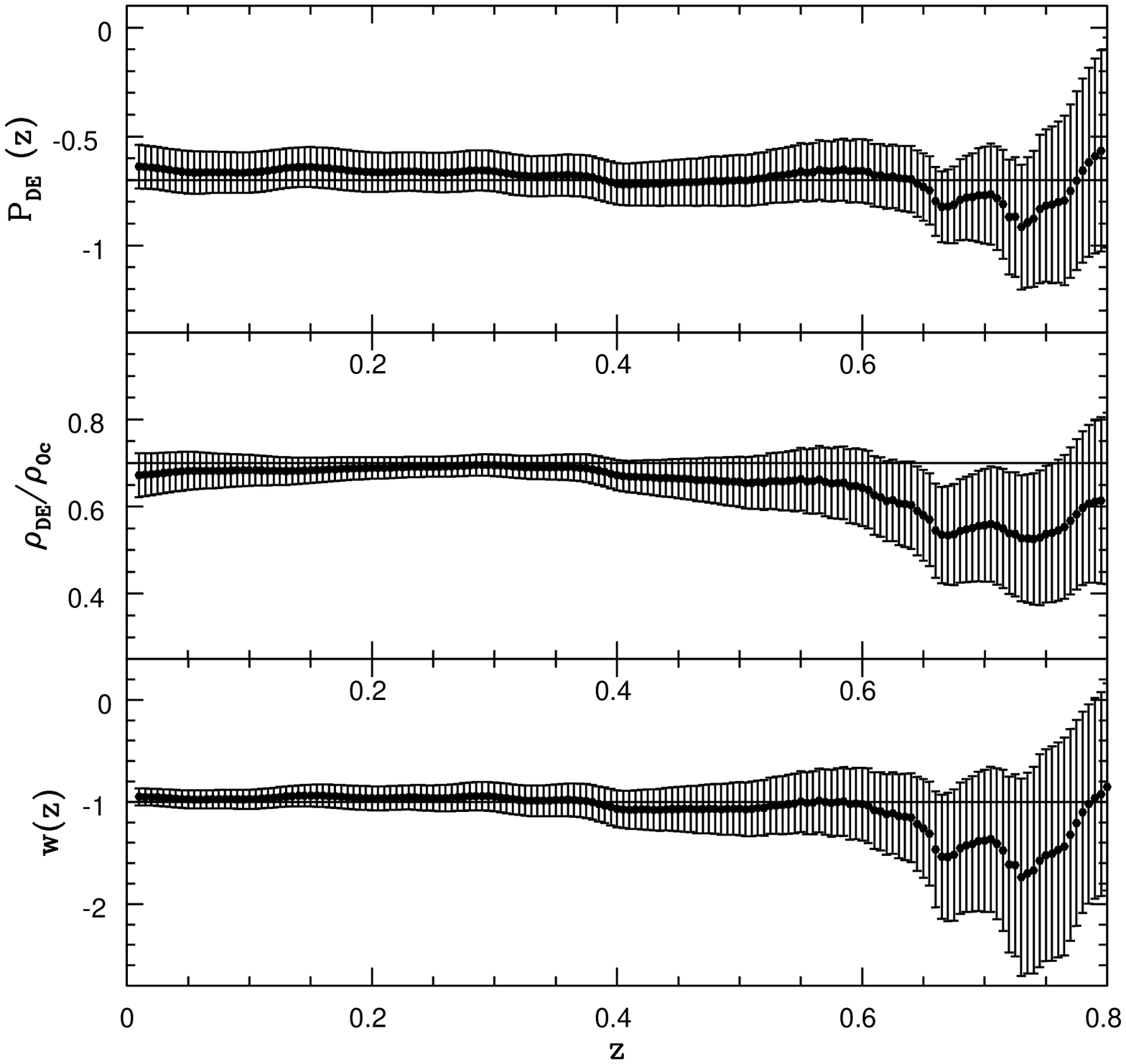}
\end{center}
\end{minipage}
\hfill
\begin{minipage}[t]{7.75cm}
\begin{center}
\includegraphics[width=7.75cm,clip]{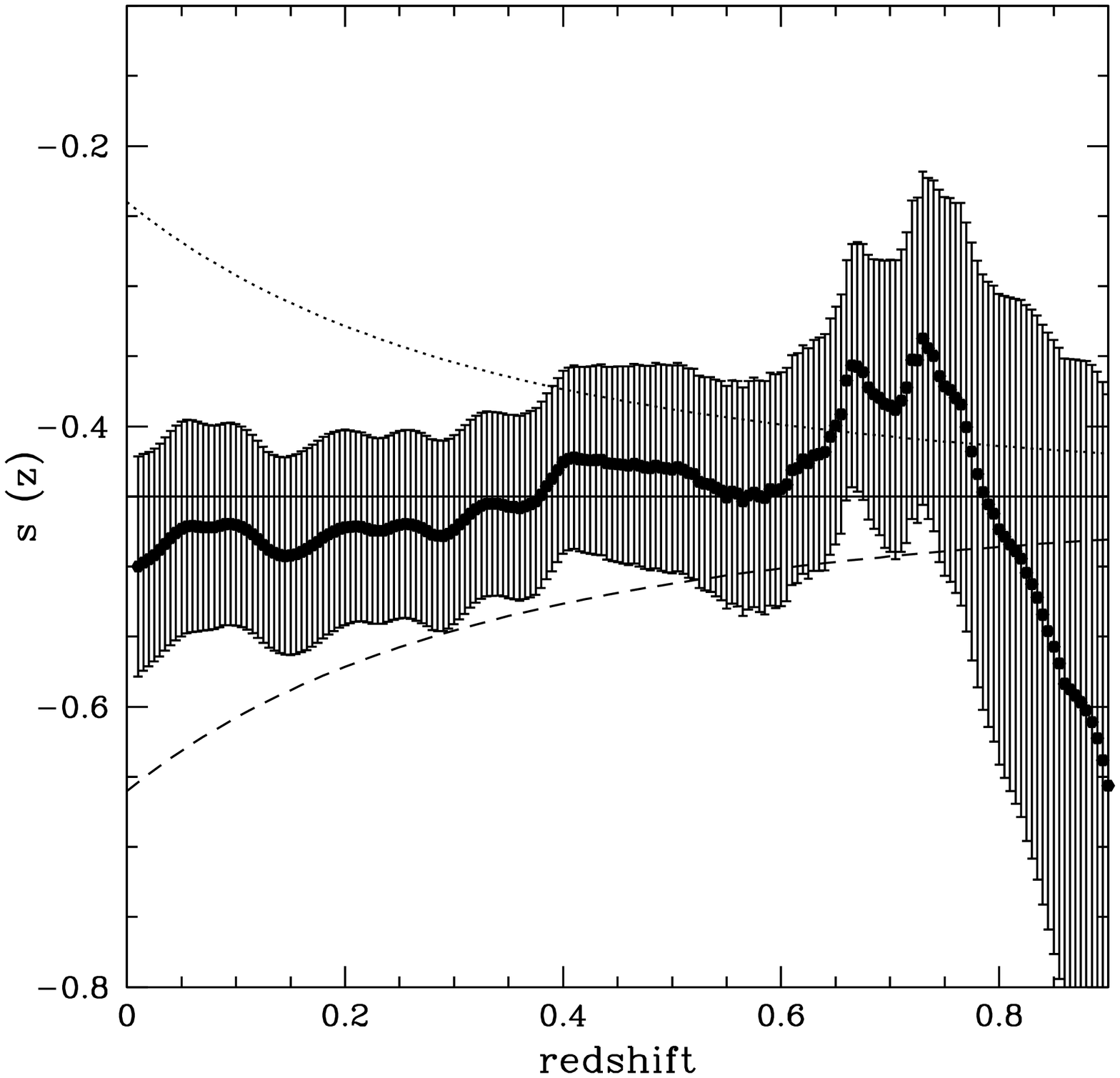}
\caption{\label{fig2}  Dark energy pressure (top left panel), 
energy density (middle left panel), and 
equation of state (bottom left panel), 
and the dark energy indicator (right panel) as functions of redshift
for a combined sample of supernovae and radio galaxies; from Daly et al. (2008).
}
\end{center}
\end{minipage}
\end{figure}


It is a pleasure to thank David Cline for encouraging me to give this
presentation, and my collaborators in these endeavors, especially 
George Djorgovski, Chris O'Dea, Preeti Kharb, and Stefi Baum. 
This work was supported in part by U.S. NSF grants AST-0096077, 
AST-0206002, and AST-0507465. 


\end{document}